\documentclass[fleqn,usenatbib]{mnras}

\usepackage{newtxtext,newtxmath}

\usepackage[T1]{fontenc}

\DeclareRobustCommand{\VAN}[3]{#2}
\let\VANthebibliography\thebibliography
\def\thebibliography{\DeclareRobustCommand{\VAN}[3]{##3}\VANthebibliography}

\usepackage{graphicx}	% Including figure files
\usepackage{amsmath}	% Advanced maths commands
\usepackage{multirow}
\usepackage{array}
\usepackage[flushleft]{threeparttable}

\title{Additional Evidence for Super-Virial Hot Phase in Milky Way CGM}
\title{Super-virial Hot Phase in Milky Way Circumgalactic Medium: Further Evidences}
\author[R. L. McClain et al.]{
Rebecca L. McClain,$^{1}$\thanks{E-mail: mcclain.378@osu.edu}
Smita Mathur,$^{1,2}$
Sanskriti Das,$^{3}$
Yair Krongold $^{4}$
and Anjali Gupta $^{5}$
\\
$^{1}$Department of Astronomy, The Ohio State University, 140 West 18th Avenue, Columbus, OH 43210, USA\\
$^{2}$Center for Cosmology and Astro-Particle Physics, The Ohio State University, 191 West Woodruff Avenue, Columbus, OH 43210, USA\\
$^{3}$Kavli Institute for Particle Astrophysics \& Cosmology, Stanford University, 452 Lomita Mall, Stanford, CA 94305, USA\\
$^{4}$Instituto de Astronomia, Universidad Nacional Autonoma de Mexico, 04510 Mexico City, Mexico\\
$^{5}$Columbus State Community College, 550 E Spring Street, Columbus, OH 43210, USA
}

\date{Accepted XXX. Received YYY; in original form ZZZ}

\pubyear{2023}

\begin{document}
\label{firstpage}
\pagerange{\pageref{firstpage}--\pageref{lastpage}}
\maketitle

\begin{abstract}
Recent discoveries of a super-virial hot phase of the Milky Way circumgalactic medium (CGM) has launched new questions regarding the multi-phase structure of the CGM around the Galaxy. We use 1.05 Ms of archival Chandra/HETG observations to characterize highly ionized metal absorption at z=0 along the line of sight of the quasar NGC\,3783. We detect two distinct temperature phases with T$_1 = 5.83^{+0.15}_{-0.07}$ K, warm-hot virial temperature, and T$_2=6.61^{+0.12}_{-0.06}$ K, hot super-virial temperature. The super-virial hot phase coexisting with the warm-hot virial phase has been detected in absorption along only two other sightlines and in one stacking analysis. There is scatter in temperature of the hot as well as warm-hot gas. Similar to previous observations, we detect super-solar abundance ratios of metals in the hot phase, with a Ne/O ratio 2$\sigma$ above solar mixtures. These new detections continue the mystery of the mechanism behind the super-virial hot phase, but provide evidence that this is a true property of the CGM rather than an isolated observation. The super-virial CGM could hold the key to understanding the physical and chemical history of the Milky Way.

\end{abstract}


%%%%%%%%%%%%%%%%%%%%%%%%%%%%%%%%%%%%%%%%%%%%%%%%%%

%%%%%%%%%%%%%%%%% BODY OF PAPER %%%%%%%%%%%%%%%%%%

\section{Introduction}
The circumgalactic medium (CGM) plays an important role in the structure and evolution of spiral galaxies, surrounding the stellar and gaseous components of the galactic disk \citep{tumlinson2017}. As the gas from the intergalactic medium (IGM) accretes onto the galaxy, it is shock heated to the virial temperature of the galaxy \citep{oppenheimer2016}. Stellar and AGN (active galactic nucleus) feedback processes expel gas from the disk of the galaxy and into the CGM, bounded by the galaxy's virial radius, or beyond to the IGM. These mechanisms enrich, excite, and mix the CGM, where the gas cools and falls back down onto the disk of the galaxy, providing new material for star formation to take place. Understanding these processes, the global structure, and the interactions between the galaxy, the CGM, and the IGM has become a topic of great discussion in observational \citep{tumlinson2017,mathur2022} and theoretical \citep{oppenheimer2016,corlies2016} work and the field has shown great development. 

Extending out to the Milky Way's virial radius of about 220 kpc, the CGM contains a considerable amount of mass and may account for the missing baryons and metals in the Galaxy \citep{gupta2012}. The majority of the matter in the CGM is shown to be in the warm-hot phase at the virial temperature of $\sim$10$^6$ K, and it is diffuse, extended, massive, and anisotropic \citep{snowden2000, henley2010, henley2013, gupta2012, gupta2014, gupta2017, nicastro2016, gautzz2018, kaaret2020}.

More recent discoveries revealed a hotter, super-virial temperature phase along several different lines of sight in both X-ray absorption \citep{das2019a-absorption,das2021,lara2023}, and X-ray emission \citep{das2019b-emission, bhattacharyya2022,bluem2022,gupta2021,gupta2023} with temperatures up to $\sim$$10^7$ K. Simultaneous analysis in emission and absorption yield a multi-phase model with temperature disagreement toward the same direction \citep{das2019a-absorption, das2019b-emission, das2021, bhattacharyya2022}. This indicates that absorption and emission spectroscopy probe different regions of the CGM, with emission probing denser regions and absorption probing more diffuse gas. Extensive all-sky emission analysis has shown that the hot phase is universal, seen all across the Galaxy \citep{bluem2022,gupta2023}. However, we cannot assume that the ubiquity of this phase in emission can be similarly applied to absorption. The super-virial hot phase has only been detected in absorption along two individual sightlines \citep{das2019a-absorption, das2021} and one stacking analysis \citep{lara2023}. The stack of observations indicates a universality of the super-virial phase, but we do not yet have enough information to claim the true distribution of the highly-ionized CGM. Thus, we aim to answer this question by adding to the collection of absorption observations in a new direction and drawing a more clear picture for the homogeneity or anisotropy of the super-virial hot phase.

In this paper we analyze CGM absorption towards NGC\,3783, a bright quasar away from the direction of the Galactic Center ($l=287.46^\circ$, $b=+22.95^\circ$). The previous single-sightline CGM absorption papers \citep{das2019a-absorption, das2021} used featureless blazars as their background continuum, but NGC\,3783 is a Seyfert galaxy meaning that is has a more complicated spectrum of its own \citep{krongold2003}. Due to the source's low redshift of z=0.0097, the intrinsic warm absorbers (WA) features are very close to the CGM absorption lines that we are interested in observing. The signal from the WAs is significantly stronger and broader than the CGM signal, making it challenging to separate the two. However, the spectral resolution of Chandra gratings is sufficiently high such that we can distinguish the WA features from the z=0 lines, allowing us to probe the Milky Way CGM.

Section \ref{data} of this paper outlines the data analysis of grating spectra from Chandra and XMM-Newton, revealing the multi-phase properties of the highly ionized CGM along this line of sight toward NGC\,3783. Section \ref{model-independent} outlines our initial motivation for the two-phase model used in Section \ref{phase-modeling}, which describes the spectral modeling conducted to determine the physical and chemical properties of the distinct phases. Section \ref{discussion} discusses the implications of our analysis, the remaining open questions, and a prediction for the upcoming XRISM mission. We finally summarize our results in Section \ref{conclusions}.

\section{Data Reduction and Analysis} \label{data}
\subsection{Chandra} \label{data-chandra}
NGC\,3783 is a bright quasar that has been repeatedly observed by X-ray telescopes. We analyze the archival Chandra data to study the z=0 Milky Way CGM absorption of the quasar emission in the wavelength range of 4-22 \AA, probing absorption lines of H-like and He-like ions of sulfur (S), silicon (Si), magnesium (Mg), neon (Ne), and oxygen (O). 

\subsubsection{Chandra Data Reduction}
We extract the data necessary for this analysis from the Chandra public data archive, taken with the high energy transmission grating (HETG) spectra on ACIS-S. Of the ten available ACIS-HETG archived observations of NGC\,3783, we choose to include eight\footnote{ObsID: 373, 2090, 2091, 2092, 2093, 2094, 14991, 15626}. The two omitted observations were taken while NGC\,3783 was obscured in 2016 \citep{kaastra2018}, decreasing the signal-to-noise (S/N) to a value below the required level for this analysis. 

We follow the standard data reduction tools and threads for grating spectra in CIAO\footnote{https://asc.harvard.edu/ciao/} (Chandra Interactive Analysis of Observations). This produces a source spectrum, background spectrum, response matrix (RMF) and auxiliary response file (ARF) for each ObsID used. We consider only the first-order spectrum of the medium-energy grating (MEG) arm of HETG, combining the positive and the negative first order spectra of all ObsIDs using the command \textit{combine\_grating\_spectra}.
We then combine the resulting positive stack and negative stack spectra to obtain one spectrum representing the sum of the first-order spectra of eight observations. The total exposure time for these observations is 1.05 Ms.

All relevant spectral analysis is conducted with Xspec in Heasoft 6.30.1\footnote{https://heasarc.gsfc.nasa.gov/docs/software/heasoft/}.

\subsubsection{Line Fitting}
The Chandra observations of NGC\,3783 span 13 years of observation, making global continuum fitting difficult due to the flux variations and the presence of the warm absorbers in the source over that time. Instead, we simplify our analysis by fitting a local continuum around each absorption line of interest, listed in Table \ref{abs-lines}, within a region of about $\pm$0.25 \AA\ around the expected central wavelength. Each local continuum is fit with a simple power law, with the normalization and the photon index as free parameters. We do not include a commonly used neutral absorber from the disk of the Galaxy (\textit{TBabs} in Xspec) in our model because there is no significant contribution to the continuum within the small range of wavelengths we consider. We place no constraints on either power law parameter to begin with, allowing Xspec to choose the best values. The presence of a strong absorption line pulls the continuum down to nonphysical levels when all parameters are free to vary. Thus we fix the photon index to the best fit value while the normalization is adjusted by eye to fit the local continuum more accurately. 

After determining the continuum parameters, we add a Gaussian absorption profile to the model with a negative normalization. We set a variable central wavelength for each of the lines, within a small range of possible values due to the presence of nearby strong quasar warm absorber features. We also fix the Gaussian width to be $10^{-5}$ \AA\ because these lines are expected to be unresolved by Chandra/HETG. 

\begin{figure}
    \centering
    \includegraphics[width=\linewidth]{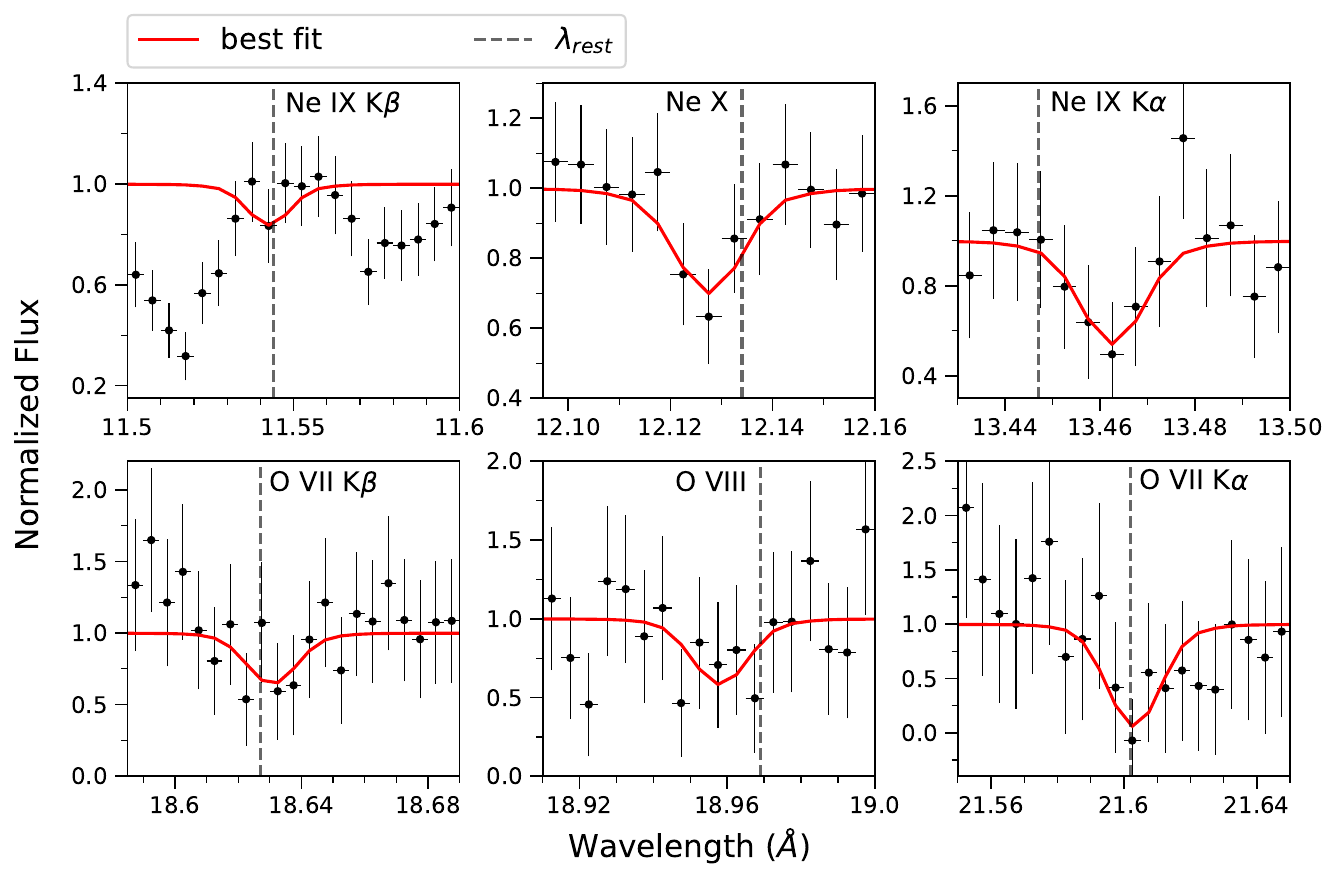}
    \caption{Best fit Gaussian profiles for the Chandra/HETG data for each of the detected z $\approx 0$ absorption lines. All spectra are folded with the line spread function of Chandra/HETG and normalized by the best-fitted local continuum. Several of the lines are redshifted or blueshifted from their expected rest wavelengths, $\lambda_{\text{rest}}$ (vertical dashed line), but the offset is consistent with $\lambda_{\text{rest}}$ within one resolution element. \ion{Ne}{IX} K$\beta$ is detected, but is contaminated by a Fe L-shell WA feature at $\sim$$\lambda=11.51$, making it difficult to model in our later analysis. The equivalent widths and implied column densities of each of the lines are shown in Table \ref{abs-lines}.}
    \label{chandra-fits}
\end{figure}

The resulting best fit equivalent width (EW) of each absorption line is shown in Table \ref{abs-lines} and the Gaussian profiles are shown in Figure \ref{chandra-fits}. Several of the absorption lines are noticeably shifted from their expected wavelength, particularly \ion{Ne}{IX} K$\alpha$ and \ion{O}{VIII}, shown by the vertical dashed line in each panel in Figure \ref{chandra-fits}. We assume this to be insignificant due to the fact that the observed wavelength shift is within $\sim$1 MEG wavelength resolution element of 23 m\AA.

We detect z=0 absorption lines of \ion{O}{VII} K$\alpha$, \ion{O}{VII} K$\beta$, \ion{O}{VIII} K$\alpha$, \ion{Ne}{IX} K$\alpha$, and \ion{Ne}{X} K$\alpha$ with 4.2$\sigma$, 2.3$\sigma$, 2.5$\sigma$, 4.1$\sigma$, and 4.5$\sigma$ significance respectively. We also detect \ion{Ne}{IX} K$\beta$ with 2.7$\sigma$ significance, but there is some contamination from Fe L-shell features from the intrinsic WA within 0.04 \AA\ of the central \ion{Ne}{IX} K$\beta$ wavelength. For this reason, we choose not to include it in our subsequent analysis. However, our detection is included in Figure \ref{chandra-fits}. We provide 3$\sigma$ upper limits on the EW of \ion{S}{XVI} K$\alpha$, \ion{Si}{XIV} K$\alpha$, \ion{Si}{XIII} K$\alpha$, \ion{Mg}{XII} K$\alpha$, and \ion{Mg}{XI} K$\alpha$, since they were not significantly detected in this data set (<2$\sigma$).
This high-significance detection of \ion{Ne}{X} is strong evidence in itself for the presence of super-virial hot gas.

\begin{table*}
    \setlength\extrarowheight{2pt}
    \centering
    \caption{Observed ions in the CGM with their corresponding rest wavelength $\lambda_{\text{rest}}$, rest energy E, oscillator strength f$_{jk}$, fit wavelength $\lambda_{\text{obs}}$, measured equivalent width EW, and associated column density N. EW$_{\text{T1}}$ and EW$_{\text{T2}}$ denote the equivalent width contributions of the lower and higher temperature gas phase respectively, determined by the non-solar two-component PHASE model. The errors quoted are 1$\sigma$ uncertainties for the parameter. 3$\sigma$ upper limits are quoted for the lines with less than 2$\sigma$ detections.}
    \label{abs-lines}
    \begin{tabular}{cccccccccc}
% \tabletypesize{\footnotesize}
        \hline
        \hline
        Ion & Transition & $\lambda_{\text{rest}}$ ({\AA}) & E (keV) & f$_{jk}$ & $\lambda_{\text{obs}}$ ({\AA}) & EW (m{\AA}) & N $\left(\times 10^{15} \text{ cm}^{-2}\right)$ & EW$_{\text{T}1}$ (m{\AA}) & EW$_{\text{T}2}$ (m{\AA})\\ 
        \hline
        \ion{S}{XVI} & K$\alpha$ & 4.729 & 2.62 & 0.274 & 4.734 & < 3.70 & < 52.72 & - & - \\
	\ion{Si}{XIV} & K$\alpha$ & 6.182 & 2.01 & 0.240 &  6.180 & < 2.86 & < 35.26 & - & - \\
        \ion{Si}{XIII} & K$\alpha$ & 6.648 & 1.87 & 0.747 & 6.655 & < 2.16 & < 7.39 & - & -\\
        \ion{Mg}{XII} & K$\alpha$ & 8.419 & 1.48 & 0.276 & 8.415 & < 3.64 & < 21.02 & - & - \\
        \ion{Mg}{XI} & K$\alpha$ & 9.169 & 1.35 & 0.741 & 9.165 & < 3.83 & < 6.95 & - & -\\
        \ion{Ne}{IX} & K$\beta$ & 11.547 & 1.08 & 0.149 & 11.544 & 2.95$^{+1.71}_{-1.09}$  & 16.79$^{+9.73}_{-6.20}$ & 1.10$^{+0.44}_{-1.10}$ & 1.95$^{+0.96}_{-0.89}$ \\
        \ion{Ne}{X} & K$\alpha$ & 12.134 & 1.02 & 0.276 & 12.130 & 5.49 $\pm$ 1.21 & 15.28 $\pm$  3.37 & - & 4.29$^{+1.54}_{-1.75}$\\
        \ion{Ne}{IX} & K$\alpha$ & 13.447 & 0.92 & 0.720 & 13.464 & 8.55 $\pm$ 2.09 & $24.98^{+16.12}_{-4.54}$ & 4.56$^{+1.29}_{-1.66}$ & 7.92$^{+2.07}_{-2.69}$ \\
        \ion{O}{VII} & K$\beta$ & 18.627 & 0.67 & 0.146 & 18.630 & 7.61 $\pm$ 3.25 & 16.97 $\pm$ 7.25 & 6.93$^{+1.21}_{-1.31}$ & - \\
        \ion{O}{VIII} & K$\alpha$ & 18.969 & 0.70 & 0.278 & 18.960 & $8.55^{+3.45}_{-3.48}$ &  $9.66^{+3.90}_{-3.93}$ & - & 5.41$^{+2.25}_{-2.34}$\\
        \ion{O}{VII} & K$\alpha$ & 21.602 & 0.57 & 0.645 & 21.603 & $19.79^{+4.03}_{-4.70}$ & $24.40^{+13.50}_{-6.00}$ & 23.49$^{+5.82}_{-6.23}$ & 2.34$^{+1.34}_{-1.13}$ \\
        \hline
    \end{tabular}
\end{table*}

\begin{figure}
    \centering
    \includegraphics[width=0.8\linewidth]{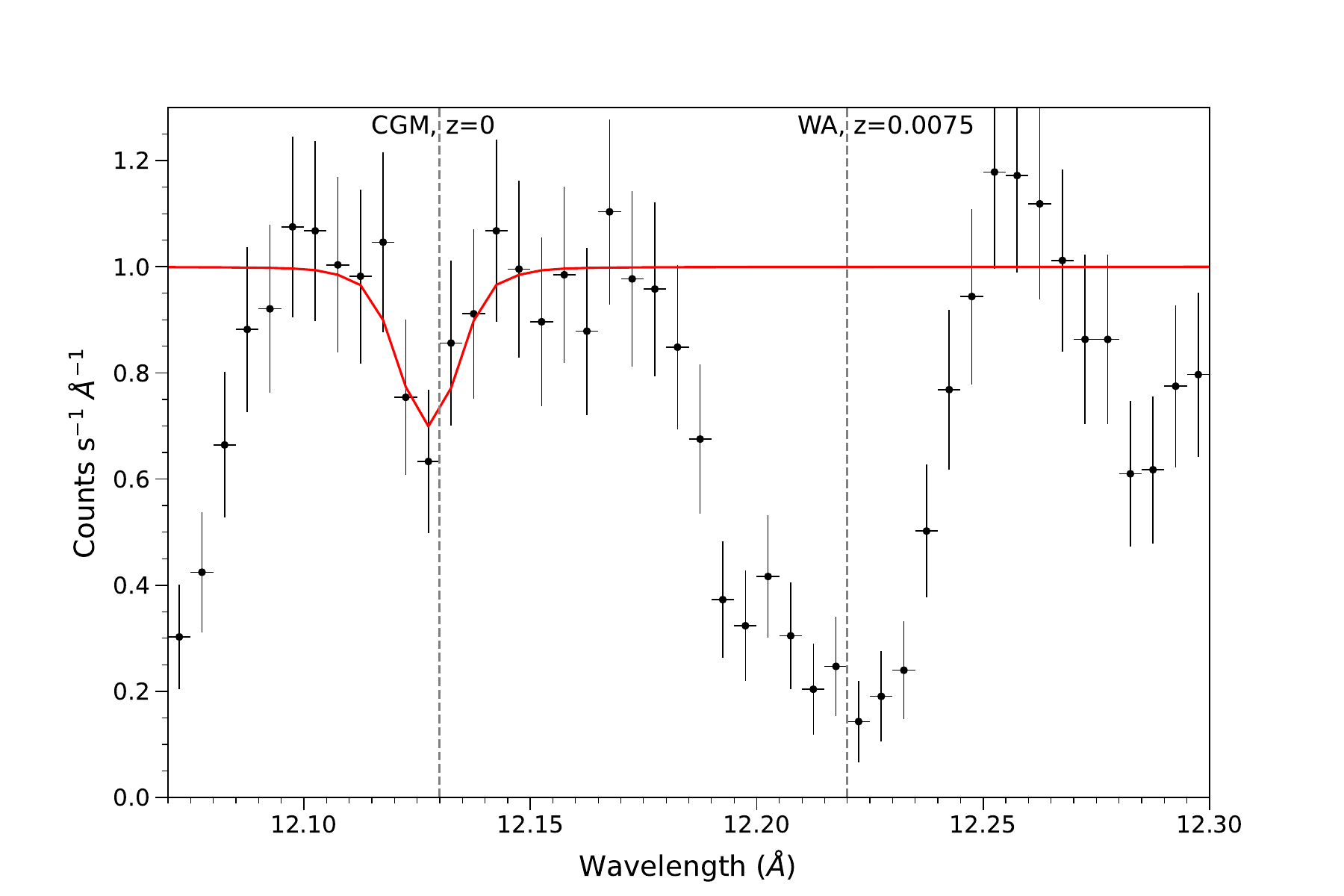}
    \caption{Chandra/HETG spectrum around the z=0 wavelength of the \ion{Ne}{X} line, showing the separation between the z=0 CGM absorption and the intrinsic z=0.0075 warm absorber (WA) feature of NGC\,3783. The z=0 \ion{Ne}{X} Gaussian profile is plotted in red. The wavelength separation between the CGM \ion{Ne}{X} line and the WA line is 0.09 \AA\ which allows us to isolate the z=0 \ion{Ne}{X} absorption in the Milky Way from the WA.}
    \label{nex-wa}
\end{figure}

We are able to isolate only the Milky Way CGM absorption for fitting and modeling due to the high spectral resolution of Chandra. Figure \ref{nex-wa} shows the separation of the z=0 CGM absorption line from the quasar warm absorber (WA) line for \ion{Ne}{X}. The source NGC\,3783 is at z=0.0097, but the \ion{Ne}{X} WA line is blueshifted from the quasar to due to the outflow velocity \citep{krongold2003}, with the observed WA redshift z=0.0075. This results in a wavelength separation of 0.09 \AA\ between the z=0 line and the intrinsic WA line. All other lines are similarly separated, allowing us to minimize the contamination from the WA features in our analysis, even for such a small redshift. Any other possible contamination from other WA lines (i.e. Fe L-shell or \ion{Ca}{XVI} transitions) is negligible, except in the case of \ion{Ne}{IX} K$\beta$ as discussed above.

\ion{O}{VII} K$\alpha$ was previously detected in \citet{gupta2012} in their analysis of the same Chandra archival data of NGC\,3783 to study the virial warm-hot gas of the Milky Way CGM. Their measured EW for this lines is $14.4\pm2.5$ m\AA\ 
while our measurement in this analysis gives an
EW of $19.79^{+4.03}_{-4.70}$. The difference in detection significance is likely due to our choice of a simple background continuum while \citet{gupta2012} chose a more complicated model to fit the continuum. However, there is still agreement between the two values within 1$\sigma$. The other lines detected in \citet{gupta2012} also shows agreement with our own measurements. \citet{fang2015} analyzed XMM-Newton observations of a collection of quasars to study \ion{O}{VII} K$\alpha$ CGM absorption, obtaining an EW of $23.30\pm6.65$ m\AA\ and $19.82\pm6.61$ m\AA\ with two different evaluation methods along the NGC\,3783 sightline. Our results remain consistent with these previous detections, providing more agreement and confidence in our new results.

\subsection{XMM-Newton}
We briefly look at the XMM-Newton Reflection Grating Spectrometer (RGS) observations of NGC\,3783 from the XMM-Newton Science Archive by stacking the observations of the unobscured quasar with a sufficient exposure time for the best possible S/N\footnote{Obs ID: 0112210101, 0112210201, 112210501}. We reduce the raw data with XMM-Newton Scientific Analysis System (SAS) and stack the resulting RGS1 and RGS2 spectra using the command \textit{rgscombine.} The resulting combined observations have a total exposure time of 316 ks. 

\begin{figure}
    \centering
    \includegraphics[width=\linewidth]{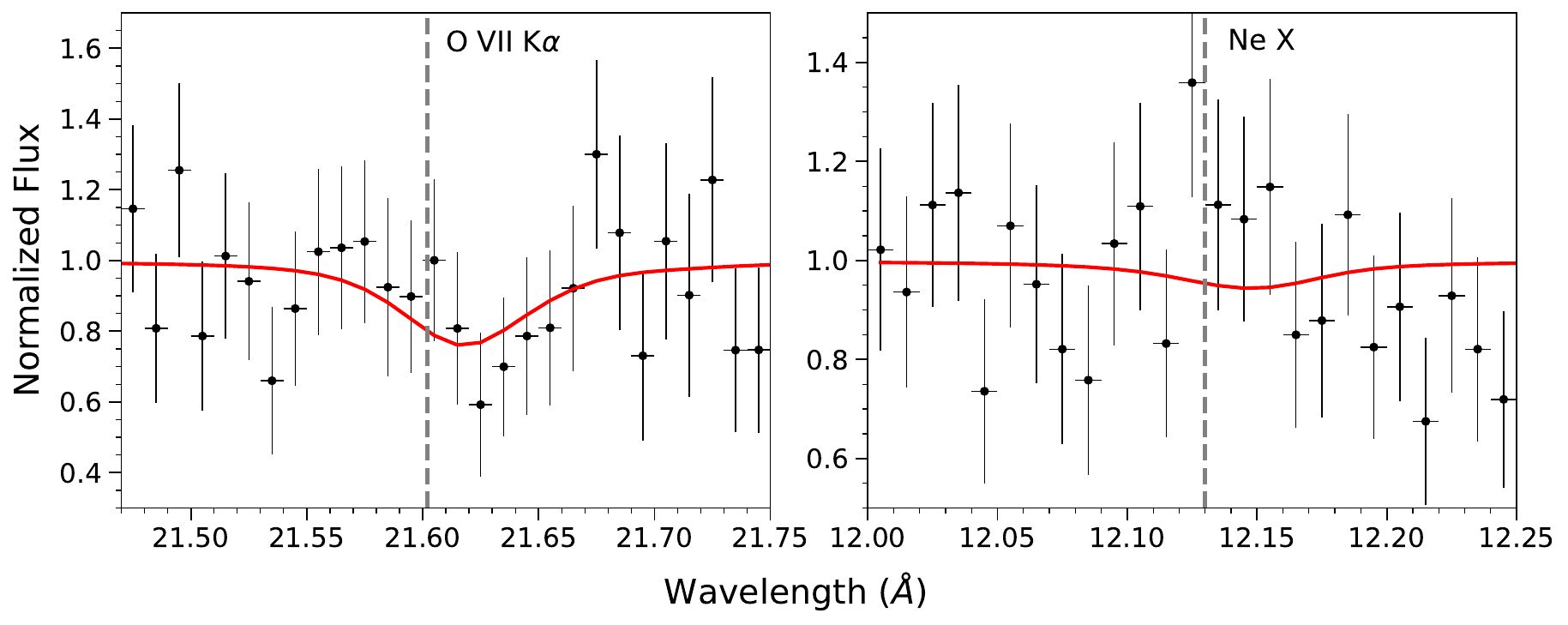}
    \caption{Normalized XMM-Newton spectra for \ion{O}{VII} K$\alpha$ (left) and \ion{Ne}{X} (right) with the best model plotted in red. The \ion{O}{VII} K$\alpha$ model is an independent fit to the XMM-Newton data, with a power law + Gaussian (centered at 21.72 \AA) continuum consistent with \citet{fang2015}. The \ion{Ne}{X} model is the best fit Chandra model in Figure \ref{chandra-fits} with an adjusted continuum normalization. The S/N in the XMM-Newton data is significantly lower than the Chandra data, so we do not expect an independent measurement of \ion{Ne}{X}, or other absorption lines. However, the \ion{O}{VII} K$\alpha$ Chandra model is consistent with the XMM-Newton data and the previous measurements.}
    \label{xmm-fits}
\end{figure}

We fit the \ion{O}{VII} K$\alpha$ line with a two-Gaussian model, one at the \ion{O}{VII} K$\alpha$  wavelength, and one centered at 21.72 \AA, consistent with the model in \citet{fang2015} with
EW = 23.81$^{+7.21}_{-6.26}$ m\AA. The spectrum around the \ion{O}{VII} K$\alpha$ line is shown in the left panel of Figure \ref{xmm-fits}, normalized by the power law + Gaussian continuum. The S/N is still too low to make any new detections, so we instead overlay our Chandra best fit on the XMM-Newton data for the \ion{Ne}{X} line, shown in the right panel of Figure \ref{xmm-fits}. This model has all the same parameters as the Chandra model, but with the power law normalization adjusted to account for continuum variability between the two observations. We see that the \ion{Ne}{X} line strength in XMM-Newton is similar to that in the Chandra observations.

Although we were unable to confirm our other detections or fit any new absorption lines, there is agreement between the two data sets in the existence of highly ionized metal absorption in the CGM.

\section{Model-Independent Results} \label{model-independent}
With our detections of highly ionized Ne and O absorption, we want to estimate the temperature of the gas containing these metals in comparison with the known warm-hot virial phase of the Milky Way CGM. The temperature can be determined based on the EW and column densities (N) measured from the spectroscopic analysis in Section \ref{data}.

We use the linear part of the curve of growth to determine the column densities of all of the unsaturated detected lines, given by Equation \ref{curve-of-growth}. 

\begin{equation}
    % EW_i /\lambda= 8.85\times10^{-13}N_i\lambda f_{jk}
    N_i = 1.13\times10^{12}\frac{EW_i}{\lambda_{jk}f_{jk}}
    \label{curve-of-growth}
\end{equation}

\ 

\hspace{-0.32cm}where $N_i$ is the column density of the ion in cm$^{-2}$, $EW_i$ is the equivalent width of the $i$-th ion in cm, $f_{jk}$ is the dimensionless oscillator strength of the given transition from the $j$-th to $k$-th atomic level, and $\lambda_{jk}$ is the wavelength of the absorption line in cm. The column densities calculated in this manner for each absorption line are shown in Table \ref{abs-lines}. 

Based on our measured EWs for \ion{O}{VII} and \ion{Ne}{IX} lines, we see that \ion{O}{VII} K$\alpha$ and \ion{Ne}{IX} K$\alpha$ are both saturated. When calculating the column densities for these lines, we combine the information from the K$\alpha$ and K$\beta$ lines as previously done in \citet{williams2005}, \citet{gupta2012}, \citet{das2021}, and \citet{mathur2022}. These values are quoted in Table \ref{abs-lines}. This method gives column densities larger than would be calculated with the linear curve of growth analysis.

In combination with the ionization fractions of each of the ions as a function of temperature ($f_{\text{ion}}(\text{T})$, taken from CHIANTI Atomic Database\footnote{https://www.chiantidatabase.org/} \citep{dere1997, delzanna2021}), we determine the approximate temperatures of the gas containing these metals. This can only be done for the cases where we have detections of two ions of the same species (\ion{O}{VII}, \ion{O}{VIII}, \ion{Ne}{IX}, and \ion{Ne}{X}). The ratio of the ionization fractions of the same species as a function of temperature is equal to the ratio of the column densities previously calculated. For example:
\begin{equation}
    \frac{f_{\ion{Ne}{IX}}(\text{T})}{f_{\ion{Ne}{X}}(\text{T})} = \frac{\text{N}_{\ion{Ne}{IX}}}{\text{N}_{\ion{Ne}{X}}}
    \label{ionfrac-column}
\end{equation}

\

For simplicity, we assume that all of the \ion{Ne}{IX} and \ion{Ne}{X} are produced by one temperature component while all of the \ion{O}{VII} and \ion{O}{VIII} are produced by another. Previous observations show that the ions are mixed between phases \citep{das2019a-absorption, das2021}, but we follow this simple assumption solely for the purpose of informing our formal model in Section \ref{phase-modeling}. 

Under this assumption, we calculate the approximate temperatures of this two-phase CGM model. The \ion{O}{VII}/\ion{O}{VIII} ratio implies log(T$_1) = 6.23\pm0.05$ K. The \ion{Ne}{IX}/\ion{Ne}{X} ratio implies log(T$_2) = 6.55^{+0.04}_{-0.05}$ K. Because there is no 1$\sigma$ overlap in temperatures of these phases, we see that there are indeed two distinct phases along this line of sight, one near the virial temperature (\ion{O}{VII}/\ion{O}{VIII}) and one at super-virial temperature (\ion{Ne}{IX}/\ion{Ne}{X}). Without modeling, there is no way to know how much each temperature component contributes to the overall EW of the lines, but this estimation gives us enough evidence to conclude that two distinct phases of the CGM are necessary to simultaneously produce the observed amount of metal absorption. 

With these model-independent temperature calculations, we can estimate the total column density of both \ion{Ne}{IX} and \ion{O}{VII}, which gives an absolute abundance ratio estimation of Ne/O = $1.24^{+1.10}_{-0.52}$, compared to a solar ratio of Ne/O = $0.23\pm 0.03$. This implies the existence of Ne abundances up to 2$\sigma$ above the solar prescriptions of \citet{asplund2021} along the sightline of NGC\,3783.

\section{PHASE Modeling} \label{phase-modeling}
We use the hybrid-ionization model PHASE (models of
collisionally ionized gas perturbed by photoionization by the
meta-galactic radiation field, at a given redshift; \citet{krongold2003}, \citet{das2019b-emission}, \cite{das2021}) to model the detected absorption lines in Chandra. As discussed above, we do not include the \ion{Ne}{IX} K$\beta$ line because the proximity and potential contamination of the Fe line does not allow for accurate modeling. The parameters in PHASE are temperature T, hydrogen column density N$_{\text{H}}$, redshift, non-thermal velocity, ionization parameter, and the abundances relative to solar of various metals (including Ne and O, detected in our analysis). We set the ionization parameter $U$ in all of our models to be $U=10^{-3.9}$, the lowest value allowed by PHASE. At this value, the photoionization is negligible, and the model is effectively a collisional ionization model only, as expected for the CGM. The non-thermal velocity is set to be zero for all absorption lines and the redshift is set based on the existing wavelength offset of the Gaussian best fit for each of the lines as discussed in Section \ref{data-chandra}.

We choose a two-component PHASE model with a power law continuum (powerlaw \textasteriskcentered{} PHASE$_1$ \textasteriskcentered{} PHASE$_2$) to account for the different temperature phases determined above. Due to the low redshift of NGC\,3783, the quasar warm absorbers have a small wavelength separation from the CGM absorption lines, creating added difficulty in a global PHASE fit. Additionally, since there is no appropriate global continuum fit used in this analysis as discussed above, we limit our fitted wavelength range to be small such that only the z=0 absorption line itself is included. This completely removes the effects of the warm absorbers and other emission or absorption lines that are beyond the focus of this analysis.  We set the local continuum levels to be the same as the chosen parameters in the Gaussian best fit, ultimately only allowing N$_{\text{H}}$, T, and Ne abundance to vary in PHASE. 

\subsection{Solar Relative Abundances}
In the first fit, we fix the relative abundances of all metals in both of the defined PHASE components to be solar, ab(X) = 1 for all elements X. We call this our "solar" model. The only metals that are detected in our analysis are O and Ne, but PHASE considers the non-detection of the included S, Si, and Mg lines in the fitting process. Each of the lines are fit simultaneously with interdependent variable parameters. Only the local continuum and redshift are unique to the line. 

The resulting best-fitted parameters are listed in Table \ref{phase-params}, with a final fit $\chi^2/$dof = $81.17/121$. Most notable is the determined temperatures of log(T$)_1 = 5.83^{+0.15}_{-0.07}$ K and log(T$)_2 = 6.61^{+0.12}_{-0.06}$ K, consistent with a two-phase CGM model with a virial and super-virial phase, as expected. Including only PHASE$_1$, $\chi^2/$dof = $96.61/121$, and including only PHASE$_2$, $\chi^2/$dof = $103.50/121$, showing that a two-temperature component model is statistically the best fit to the data. With this model, PHASE fits the \ion{O}{} lines appropriately, giving EWs consistent with our Gaussian fits. However, the model fits the \ion{Ne}{X} absorption line poorly, with an EW of 1.85$^{+0.55}_{-0.83}$ m\AA, underestimating the strength of the \ion{Ne}{X} line by about 3$\sigma$ from the Gaussian model. This difference indicates a non-solar mixture of metals, since even with a higher temperature, PHASE cannot replicate the amount of \ion{Ne}{X} present under the assumption of solar abundance ratios. The full solar fit is shown with a dotted line in Figure \ref{phase_comp_fit}. Note the underproduction of \ion{Ne}{X} in the top left panel.

Because the \ion{Ne}{X} is underestimated in PHASE$_2$ with a solar model, our model suggests a non-solar Ne/O abundance ratio to account for the differences in Ne and O absorption. A variation in any other PHASE parameters is not sufficient.

\subsection{Non-Solar Relative Abundances}
Allowing super-solar abundance ratios, as necessitated by the results of the solar model, gives the "non-solar" model quoted in Table \ref{phase-params}. We freeze the temperatures at their best-fitted values in the solar model, instead varying the metal abundances for both PHASE components. The ionization parameter, redshift, and non-thermal velocity are still fixed to be as defined in the solar model, and N$_{\text{H}}$ remains variable. 

PHASE assumes a solar absolute metallicity when determining the value of N$_{\text{H}}$, but we do not measure a true metallicity along this line of sight. Therefore, we refer to all metal abundances as relative ratios. In our model, the Ne abundance varies relative to O, which is held fixed at solar metallicity. The O lines were fit well in our solar model, so we take this as a good approximation. We do not have enough information to be able to vary both ab(Ne) and ab(O) relative to solar in PHASE as these are the only detected metals.

The resulting best fit has $\chi^2/$dof = $75.48/121$, a significant improvement over the solar model. PHASE$_1$, dependent on \ion{Ne}{IX}, requires no change in \ion{Ne}{}abundance, fit well with ab(Ne)$_1 = 1.00^{+2.49}$. We fixed the abundance to be solar or above, so there is no 1$\sigma$ lower limit on this value. However, PHASE$_2$, defined largely by the overabundant \ion{Ne}{X}, requires ab(Ne)$_2 = 5.51^{+7.47}_{-2.84}$ (or $[\text{Ne/O]} = 0.75^{+0.38}_{-0.30}$ following \citet{asplund2021}). This is $\sim$2$\sigma$ above solar abundances. When we force Ne/O to be the same between the two fit components, we get ab(Ne) $=3.59^{+3.16}_{-1.64},$ consistent within 1$\sigma$ with our best fit.
There must be super-solar Ne abundances in the super-virial hot phase of the CGM along the line of sight of NGC\,3783. 

Allowing N$_{\text{H}}$, T, and ab(Ne) to vary simultaneously from our best fit values results in parameters within 1$\sigma$ of those quoted in Table \ref{phase-params}. Thus our choice to freeze the temperatures in PHASE$_2$ does not impact our results with significance.  

\begin{figure*}
    \centering
    \includegraphics[width=\textwidth]{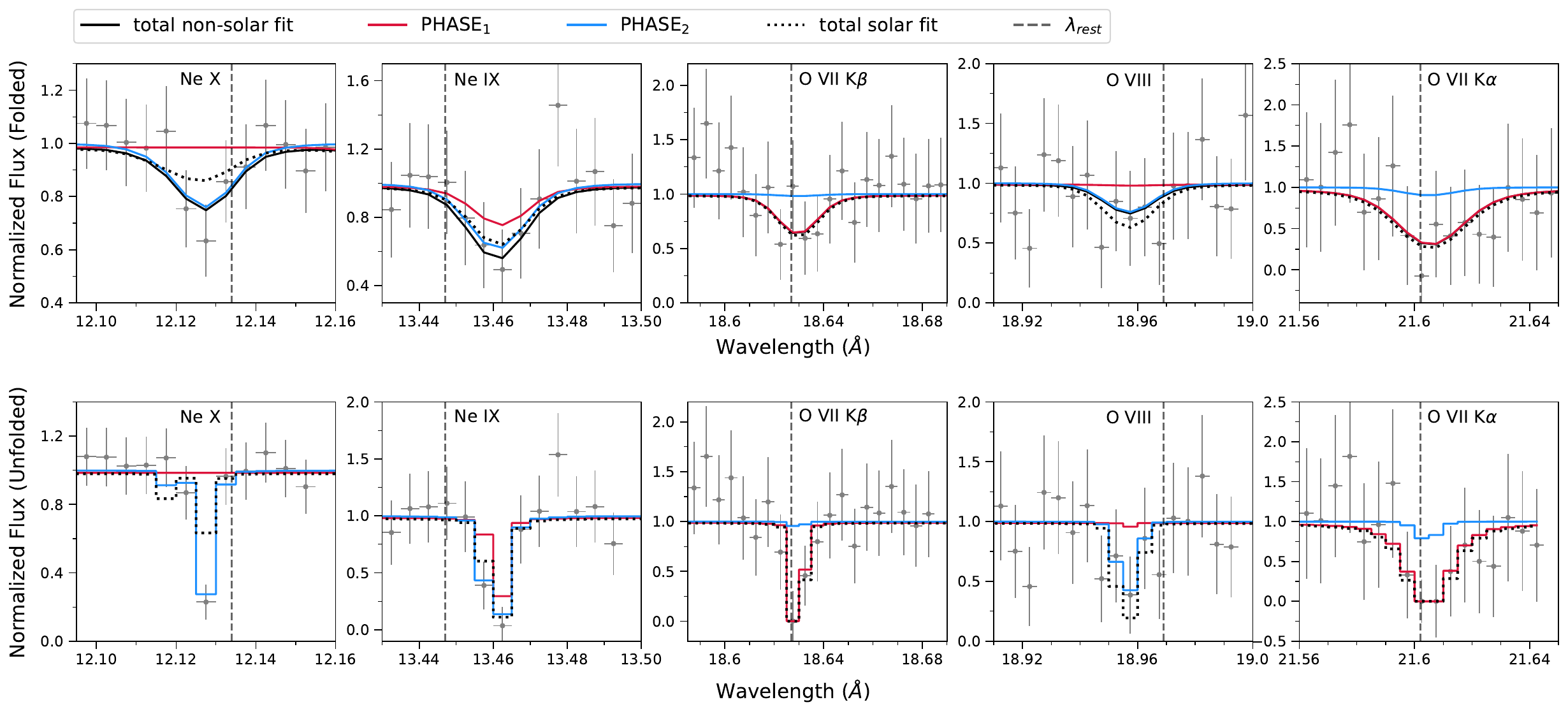}
    \caption{Non-solar PHASE models for the detected absorption lines in the Chandra spectrum. The spectrum folded with the line spread function of Chandra/HETG is shown in the top row and the unfolded spectrum is shown in the bottom row. Both spectra have been normalized by the best-fitted local continuum. The best-fitted two-T PHASE model is shown in black solid lines, with the separate contributions from the lower temperature PHASE component, PHASE$_1$ (red), and the higher temperature PHASE component, PHASE$_2$ (blue). All of the \ion{Ne}{X} and \ion{O}{VIII} is produced by the super-virial temperature component, while all of the \ion{O}{VII} is produced by the virial temperature component. The solar fit is also plotted with a dotted black line, showing that solar abundances poorly fit the \ion{Ne}{X} line, indicating the necessity for non-solar Ne/O abundance in the hot component.}
    \label{phase_comp_fit}
\end{figure*}

The non-solar model for each of the detected lines is shown in Figure \ref{phase_comp_fit} with each of the PHASE component contributions highlighted in different colors. The top panel shows the spectrum folded with the line spread function of Chandra/HETG, while the bottom panel shows the unfolded spectrum. We see that all of the \ion{Ne}{X} and \ion{O}{VIII} absorption comes from the hot phase, while almost all of the \ion{O}{VII} comes from the warm-hot phase. \ion{Ne}{IX} K$\alpha$ is split between both phases. Here, the \ion{Ne}{X} line is well fit with the high \ion{Ne}{}abundance, consistent with the Gaussian fit. The \ion{Ne}{IX} K$\beta$ line was not included in the PHASE-modeled data, but the resulting predicted EW for \ion{Ne}{IX} K$\beta$ is consistent with our Gaussian fit measurement. The two temperature component contributions to the EW are shown in Table \ref{abs-lines}. With this consistency, we can conclude that \ion{Ne}{IX} K$\beta$ is likely not contaminated by the local Fe lines, but their proximity still complicates the independent modeling.

\begin{table}
    \setlength\extrarowheight{2pt}
    \centering
    \caption{PHASE best fit model parameters and their 1$\sigma$ uncertainties for all detected CGM absorption lines with both solar and non-solar abundance ratios. The values with no quoted errors were fixed during the modeling and thus have no associated uncertainty. The EW of the resulting absorption lines of the non-solar model are shown in Table \ref{abs-lines}.} 
    \begin{threeparttable}
    \begin{tabular}{ccccc}
        \hline
        \hline
         & Parameter & & \multicolumn{2}{c}{Values} \\
         \hline
         & & & Solar & Non-Solar \\
         \hline
         \multirow{3}{4em}{PHASE$_1$} & log(N$_\text{H}$)$_1$ & & $20.20^{+0.22}_{-0.28}$ cm$^{-2}$ & $20.10_{-0.31}^{+0.22}$ cm$^{-2}$\\
         & log(T)$_1$ & & $5.83^{+0.15}_{-0.07}$ K & 5.83 K \\
         & ab(Ne)$_1$ & &  1.00 & $1.00^{+2.49}_{*}$ \\
        \hline
        \multirow{3}{4em}{PHASE$_2$} & log(N$_\text{H}$)$_2$ & & $19.92^{+0.15}_{-0.18}$ cm$^{-2}$ & $19.65^{+0.22}_{-0.31}$ cm$^{-2}$ \\
         & log(T)$_2$ & & $6.61^{+0.12}_{-0.06}$ K & 6.61 K\\
         & ab(Ne)$_2$ & & 1.00 & $5.51^{+7.47}_{-2.84}$ \\
         \hline
    \end{tabular}
    \begin{tablenotes}
    \item * The lower limit of ab(Ne)$_1$ is pegged at 1.
    \end{tablenotes}
    \end{threeparttable}
    \label{phase-params}
\end{table}

At fixed T and fixed ab(O), N$_\text{H}$ is anti-correlated with ab(Ne) in the hot phase (PHASE$_2$). The hot phase is characterized dominantly by \ion{Ne}{X} and \ion{O}{VIII}. The \ion{O}{ }lines are sensitive to the changes in N$_\text{H}$, not in changes to ab(Ne), while \ion{Ne}{} is sensitive to changes in both parameters. Leaving log(N$_\text{H}$)$_2$ and ab(Ne)$_2$ free, PHASE chooses a combination that best fits both ions. Increasing N$_\text{H}$ decreases the necessary ab(Ne) to fit \ion{Ne}{X}, but also increases the strength of the \ion{O}{VIII} line. We have no prior knowledge of the expected hydrogen column density of the super-virial hot phase, so we can make no distinction between the chosen values in the solar and non-solar model. If, however, N$_\text{H}$ is higher than our best-fit modeled value (still assuming fixed T and solar ab(O)), this would imply a lower---only $1\sigma$ super-solar---Ne/O than currently predicted. 

We find no similar correlation between the temperatures of PHASE$_1$ and PHASE$_2$ nor between the temperature and Ne/O of PHASE$_2$.

\section{Discussion} \label{discussion}
We find that two distinct temperature phases coexist in the Milky Way CGM along the line of sight of NGC\,3783 at temperatures at and above the virial temperature, with an overabundance of Ne in only the hot phase. This the fourth detection of the super-virial hot phase and the third detection of super-solar abundance ratios in absorption from the CGM. An in-depth discussion of the possible explanations for the super-virial temperatures and metal enrichment within the CGM has already been outlined in \citet{das2019a-absorption} and \citet{das2021}, thus we will only reiterate the important and relevant information here. 

\subsection{Implications for CGM Structure}

\begin{figure*}
    \centering
    \includegraphics[width=\textwidth]{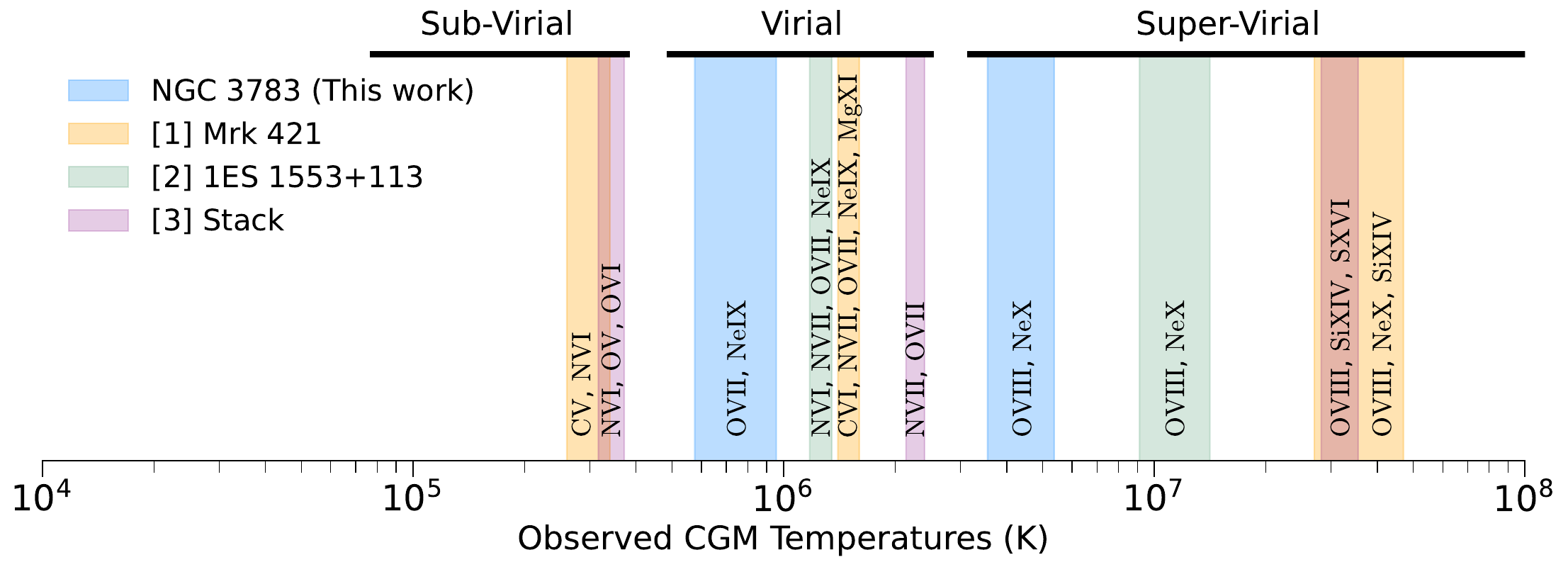}
    \caption{A summary of the current absorption detections of the warm (sub-virial), warm-hot (virial), and hot (super-virial) phases of the Milky Way CGM. The width of the band represents the 1$\sigma$ uncertainty (3$\sigma$ for Mrk 421) on the measured temperature. The coexistence of three distinct phases of the highly-ionized CGM has been observed along three individual sightlines and in a stack of 46 low S/N galaxies. There is large scatter across orders of magnitude in temperature, even within each phase. [1] \citet{das2021}, [2] \citet{das2019a-absorption}, [3] Lara-DI et al. 2023, in preparation.}
    \label{cgm-temps}
\end{figure*}

We have now detected the super-virial phase in absorption on four accounts, three individual sightlines with high S/N data and one stack of low S/N data towards other sightlines. Each of these studies has detected super-virial temperatures based on the presence of different highly ionized (H-like) metals, e.g., \ion{O}{VIII}, \ion{Ne}{X}, \ion{Si}{XIV}, \ion{S}{XVI}. Even with this now seemingly universal property, there is still significant scatter in temperature within the super-virial phase and the well-understood virial phase as depicted in Figure \ref{cgm-temps}. In these observations, we see three distinct phases in temperature---sub-virial, virial, and super-virial---with almost an order of magnitude variation in the super-virial phase and about a factor of 3 scatter in the virial phase. This temperature scatter has also been observed in emission \citep{das2019b-emission, gupta2023, bhattacharyya2022}. The consistent observation indicates that regions of gas in the CGM can be much hotter than the volume-filling virial phase, but is not homogeneously mixed or heated.

In a stack of 46 quasar sightlines, \citet{lara2023} detect \ion{Si}{XIV} and \ion{S}{XVI} associated with the super-virial hot component of the Milky Way CGM at temperatures of T $\approx10^{7.5}$ K (temperature determination described in Lara-DI et al. 2023, in preparation). However, they do not detect any \ion{Ne}{X} in their resulting spectrum, which has been necessary to constrain the hot temperature in the other studies, including this one. The main reason that \ion{Ne}{X} is detectable is because of an overabundance of Ne, increasing the strength of the faint lines. This may be indicative of inhomogeneous mixing of metals, with some regions of the CGM being more Ne-rich than others. 

The CGM of the Milky Way is not uniform, it is patchy in chemical composition and temperature. Galactic feedback and outflow processes cause heating and turbulence in the CGM. In turn, turbulence can affect the heating, cooling, and mixing of the CGM gas. What we observe is the complex effects of all of these interactions, which combine to create the multi-phase structure of the highly ionized CGM.

There are still many open questions regarding this phase-structure and the mechanisms powering the CGM. What processes are able to sustainably heat the gas to such hot temperatures observed? We see that the virial and super-virial phases coexist along the line of sight, but where does the hotter gas reside along that line? Given the likely solar Ne/O in the warm-hot component, but significantly super-solar Ne/O in the hot phase, we speculate that the two phases might not be co-spatial, and might arise from different physical processes. For example, the hot phase may occupy the extra-planar region of the Galaxy, while the warm-hot phase may fill the extended region out to the virial radius. Of course, adding more lines of sight to this group of analysis will allow us to probe more of the hot phase structure around the CGM and explore more of the metal mixtures that exist around the Galaxy. 

This super-virial phase and non-solar chemical composition is still new. However, in the absence of non-detections of the super-virial phase, we are confident in its universal presence in the Milky Way CGM. In their analysis of 3-D hydrodynamical simulations, \citet{vijayan2022} were able to reproduce the hot multi-phase gas in the CGM. They find that, in the case of high star formation rates, an asymmetric distribution of temperature and metallicity exists inside and outside the produced biconical outflow, which may relate to the wide range of temperatures (Figure \ref{cgm-temps}) and metal ratios in our observations. The temperature that we measure along the sightline of NGC\,3783 (log(T)$=6.61^{+0.12}_{-0.06}$ K) is reproduced in the extra-planar regions in \citet{vijayan2022} and is a direct result of the stellar feedback in the galaxy. The combination of theory, simulations, and observations will allow us to shed light on the processes governing the Milky Way CGM. Understanding the nature of the super-virial hot phase, and the amount of baryonic matter and metals contained in it, is important for understanding the CGM formation, and thus, galactic evolution.

\subsection{XRISM Prediction} \label{xrism-prediction}
With the upcoming XRISM mission, new doors are opening in the field of X-ray astronomy and an understanding the capabilities of the instruments in achieving our science is necessary. XRISM could be used to study the Milky Way CGM in greater depth, helping us understand the underlying mechanisms causing the observations we have obtained thus far. With the planned observations of NGC\,3783, the results of this analysis could be confirmed and expanded with the better spectral resolution and higher effective area of XRISM. We simulate XRISM/Resolve 7 eV resolution observations of NGC\,3783 based on our non-solar PHASE model to explore the potential results of new observations under the same analysis. 

\begin{figure}
    \centering
    \includegraphics[width=\linewidth]{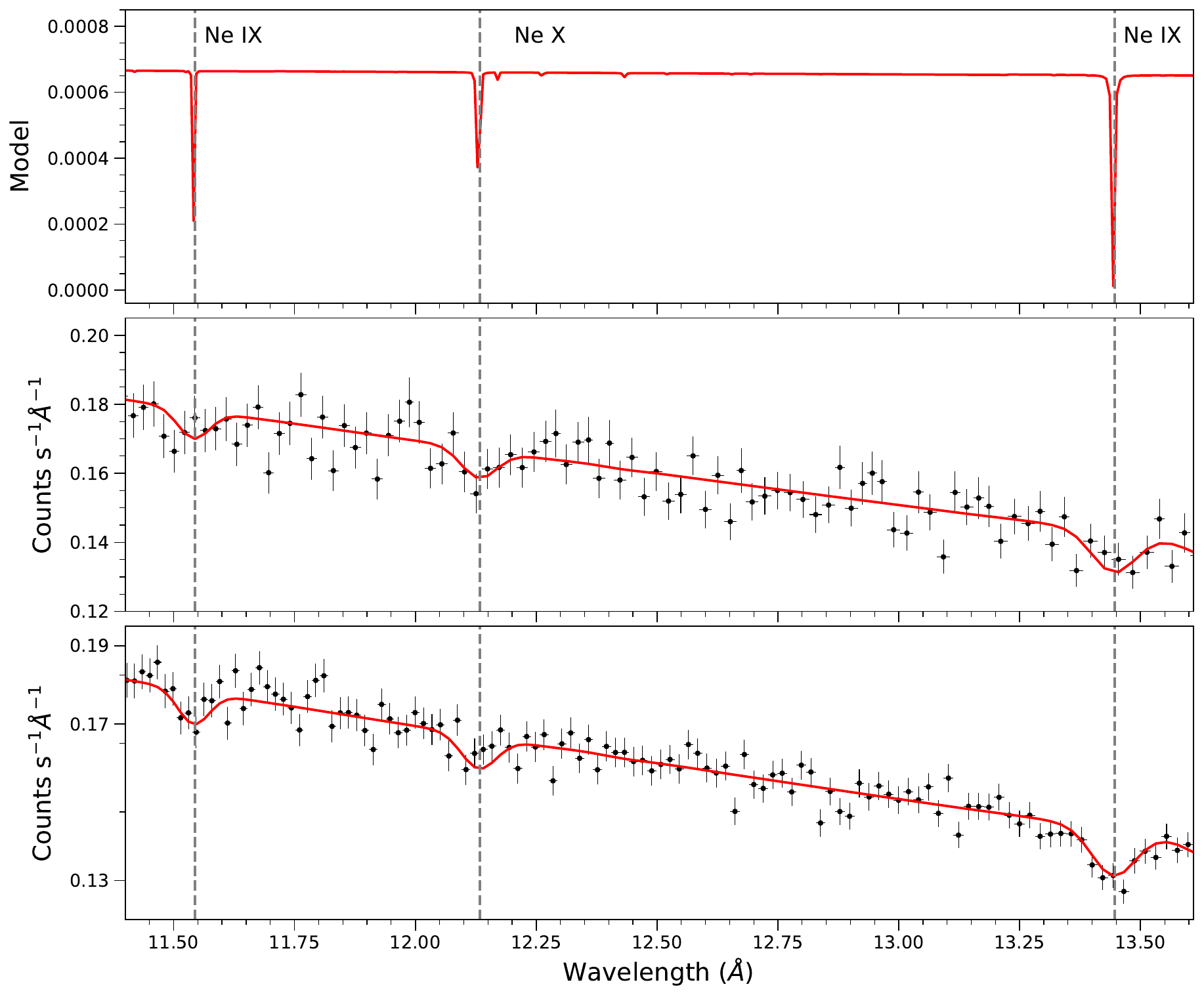}
    \caption{Simulated XRISM/Resolve 7 eV resolution observations showing predicted detections of \ion{Ne}{}lines along the NGC\,3783 line of sight. Top: The PHASE model used to simulate the XRISM spectra. Middle: 200ks integration time. Bottom: 600ks integration time, representing an observation at 3 times the flux of the middle panel. Both spectra are folded with the line spread function of XRISM/Resolve. An observation at a time when NGC\,3783 has a higher flux will increase the S/N in the data, increasing the significance of detections and allowing us to measure new lines not currently detected in Chandra data.}
    \label{xrism-sim}
\end{figure}

The source is inherently variable, meaning that the observed flux changes over time, influencing the quality of the received data. We simulate an observation with 200 ks of integration time at the measured flux of the Chandra data ($1.72^{+0.005}_{-0.003}\times10^{-11}$ erg/cm$^2$/s between 0.295 and 2.4 keV). The resulting spectrum is shown in Figure \ref{xrism-sim} (middle row). We bin the spectrum for visual effect, but the quality of the data at 200 ks is roughly the same as 1 Ms of Chandra data, with an EW limit of 1.82 m\AA.

However, at energies below about 0.7 keV (wavelengths at which the O lines can be detected), the effective area of XRISM/Resolve is below 100 cm$^2$. This means that we are not able to detect O absorption lines with the same level of significance as Chandra. The 200 ks XRISM simulation measures \ion{O}{VII} K$\alpha$ with an EW of 27.31$^{+7.41}_{-7.37}$ m\AA, which has a lower significance (3.71$\sigma$) and larger uncertainties than our Chandra measurement. Thus, at longer wavelengths, Chandra is preferred when compared to a shorter XRISM observation.

Archival ROSAT observations of NGC\,3783 obtained a flux up to $5.66\times10^{-11}$ erg/s/cm$^2$ \citep{rosatcat2018} in the range 0.1-2.4 keV, approximately 3 times higher than our Chandra observations. This would be equivalent to observing for 3 times longer, yielding a spectrum with a higher S/N and more significant detections of the absorption lines of interest. We simulate another XRISM spectrum, instead with 600 ks of exposure time to predict the results from a higher flux observation, also shown in Figure \ref{xrism-sim} (bottom row). With the longer exposure time, many of the absorption lines would be detected with more significance, which can be noticed by eye in the spectrum. This would allow for more precise measurements of line strengths, metal abundances, and temperature in the CGM.  

We suggest that NGC\,3783 be monitored over time (with Swift, for example) and observed with XRISM at a period of high flux. This would allow us to detect more of the lines predicted by PHASE and confirm the existence of the super-virial temperature gas with different metals. Our PHASE model predicts many additional lines that could be potentially detected with as little as 200 ks of integration time. All of the lines with PHASE-predicted equivalent widths within the wavelength range of XRISM/Resolve are listed in Table \ref{phase-predictions}. Detections of additional z=0 metals in the spectrum would allow us to better constrain the CGM metal mixture along this line of sight, giving us more than just \ion{Ne}{}and \ion{O}{}to add to our understanding of the thermal and chemical structure of the CGM.

\section{Conclusions} \label{conclusions}
In this paper, we present deep Chandra/HETG observations of NGC\,3783 that show z=0 metal line absorption from the Milky Way CGM. 
\begin{itemize}
    \item We fit the continuum around each of the absorption lines with a simple power law and add a Gaussian profile to determine EW of each detected absorption line. We see agreement when comparing to previous analysis of Chandra \citep{gupta2012} and XMM-Newton \citep{fang2015} data of NGC\,3783.
    \item We use PHASE to properly model the spectrum and determine the temperatures of the two detected CGM phases to be T$_1$=$5.83^{+0.15}_{-0.07}$ K, warm-hot virial temperature, and T$_2=6.61^{+0.12}_{-0.06}$ K, hot super-virial temperature.
    \item We detect a super-solar Ne abundance of ab(Ne)$_2 = 5.51^{+7.47}_{-2.84}$ relative to solar, or $[\text{Ne/O]} = 0.75^{+0.38}_{-0.30}$, which is $\sim$2$\sigma$ above solar in the hot phase, and an abundance consistent with solar in the warm-hot phase. However, if the total N$_\text{H}$ is higher than predicted, our models call for a Ne abundance closer to solar with ab(Ne)$_2$ $\approx$ 2.
    \item We make predictions for upcoming XRISM/Resolve observations and find that the same quality of data can be achieved with XRISM in only 200 ks, compared to 1 Ms with Chandra. We also suggest that, in order to maximize CGM absorption line detections, NGC\,3783 be observed in a high flux state. 
\end{itemize}

In conjunction with the previous absorption observations of the Milky Way CGM hot phase, this analysis shows that the hot super-virial phase can be seen around the Galaxy and provides confidence in the multi-phase structure of the CGM. However, as discussed in Section \ref{discussion}, there remains variation even within the super-virial phase. Determining the underlying mechanisms behind the existence of the hot phase will require more observational and theoretical work in the development of this field.

\section*{Acknowledgements}
SM is grateful for the grant provided by the National Aeronautics
and Space Administration (NASA) through \textit{Chandra} Award Number
AR0-21016X issued by the \textit{Chandra} X-ray Center, which is operated
by the Smithsonian Astrophysical Observatory, for and on behalf
of the National Aeronautics Space Administration under contract
NAS8-03060. S.M. is also grateful for the NASA ADAP grant no.
80NSSC22K1121. S.D. acknowledges support from the KIPAC Fellowship of Kavli Institute for Particle Astrophysics and Cosmology,
Stanford University. AG gratefully acknowledges support through
the NASA ADAP grant no. 80NSSC18K0419. YK acknowledges
support from UNAM PAPIIT grant no. IN102023.

%%%%%%%%%%%%%%%%%%%%%%%%%%%%%%%%%%%%%%%%%%%%%%%%%%
\section*{Data Availability}

All data used in the presented analysis is available online via the Chandra Data Archive (CDA) and the XMM-Newton Science Archive (XSA). The relevant ObsID numbers are provided below:

Chandra: 373, 2090, 2091, 2092, 2093, 2094, 14991, 15626

XMM-Newton: 0112210101, 0112210201, 112210501



\bibliographystyle{mnras}
\bibliography{references} % if your bibtex file is called example.bib

\begin{thebibliography}{}
\makeatletter
\relax
\def\mn@urlcharsother{\let\do\@makeother \do\$\do\&\do\#\do\^\do\_\do\%\do\~}
\def\mn@doi{\begingroup\mn@urlcharsother \@ifnextchar [ {\mn@doi@} {\mn@doi@[]}}
\def\mn@doi@[#1]#2{\def\@tempa{#1}\ifx\@tempa\@empty \href {http://dx.doi.org/#2} {doi:#2}\else \href {http://dx.doi.org/#2} {#1}\fi \endgroup}
\def\mn@eprint#1#2{\mn@eprint@#1:#2::\@nil}
\def\mn@eprint@arXiv#1{\href {http://arxiv.org/abs/#1} {{\tt arXiv:#1}}}
\def\mn@eprint@dblp#1{\href {http://dblp.uni-trier.de/rec/bibtex/#1.xml} {dblp:#1}}
\def\mn@eprint@#1:#2:#3:#4\@nil{\def\@tempa {#1}\def\@tempb {#2}\def\@tempc {#3}\ifx \@tempc \@empty \let \@tempc \@tempb \let \@tempb \@tempa \fi \ifx \@tempb \@empty \def\@tempb {arXiv}\fi \@ifundefined {mn@eprint@\@tempb}{\@tempb:\@tempc}{\expandafter \expandafter \csname mn@eprint@\@tempb\endcsname \expandafter{\@tempc}}}

\bibitem[\protect\citeauthoryear{{Asplund}, {Amarsi}  \& {Grevesse}}{{Asplund} et~al.}{2021}]{asplund2021}
{Asplund} M.,  {Amarsi} A.~M.,   {Grevesse} N.,  2021, \mn@doi [\aap] {10.1051/0004-6361/202140445}, \href {https://ui.adsabs.harvard.edu/abs/2021A&A...653A.141A} {653, A141}

\bibitem[\protect\citeauthoryear{{Bhattacharyya}, {Das}, {Gupta}, {Mathur}  \& {Krongold}}{{Bhattacharyya} et~al.}{2022}]{bhattacharyya2022}
{Bhattacharyya} S.,  {Das} S.,  {Gupta} A.,  {Mathur} S.,   {Krongold} Y.,  2022, \mn@doi [arXiv e-prints] {10.48550/arXiv.2208.07863}, \href {https://ui.adsabs.harvard.edu/abs/2022arXiv220807863B} {p. arXiv:2208.07863}

\bibitem[\protect\citeauthoryear{{Bluem} et~al.,}{{Bluem} et~al.}{2022}]{bluem2022}
{Bluem} J.,  et~al., 2022, \mn@doi [\apj] {10.3847/1538-4357/ac8662}, \href {https://ui.adsabs.harvard.edu/abs/2022ApJ...936...72B} {936, 72}

\bibitem[\protect\citeauthoryear{{Corlies} \& {Schiminovich}}{{Corlies} \& {Schiminovich}}{2016}]{corlies2016}
{Corlies} L.,  {Schiminovich} D.,  2016, \mn@doi [\apj] {10.3847/0004-637X/827/2/148}, \href {https://ui.adsabs.harvard.edu/abs/2016ApJ...827..148C} {827, 148}

\bibitem[\protect\citeauthoryear{{Das}, {Mathur}, {Nicastro}  \& {Krongold}}{{Das} et~al.}{2019a}]{das2019a-absorption}
{Das} S.,  {Mathur} S.,  {Nicastro} F.,   {Krongold} Y.,  2019a, \mn@doi [\apjl] {10.3847/2041-8213/ab3b09}, \href {https://ui.adsabs.harvard.edu/abs/2019ApJ...882L..23D} {882, L23}

\bibitem[\protect\citeauthoryear{{Das}, {Mathur}, {Gupta}, {Nicastro}  \& {Krongold}}{{Das} et~al.}{2019b}]{das2019b-emission}
{Das} S.,  {Mathur} S.,  {Gupta} A.,  {Nicastro} F.,   {Krongold} Y.,  2019b, \mn@doi [\apj] {10.3847/1538-4357/ab5846}, \href {https://ui.adsabs.harvard.edu/abs/2019ApJ...887..257D} {887, 257}

\bibitem[\protect\citeauthoryear{{Das}, {Mathur}, {Gupta}  \& {Krongold}}{{Das} et~al.}{2021}]{das2021}
{Das} S.,  {Mathur} S.,  {Gupta} A.,   {Krongold} Y.,  2021, \mn@doi [\apj] {10.3847/1538-4357/ac0e8e}, \href {https://ui.adsabs.harvard.edu/abs/2021ApJ...918...83D} {918, 83}

\bibitem[\protect\citeauthoryear{{Del Zanna}, {Dere}, {Young}  \& {Landi}}{{Del Zanna} et~al.}{2021}]{delzanna2021}
{Del Zanna} G.,  {Dere} K.~P.,  {Young} P.~R.,   {Landi} E.,  2021, \mn@doi [\apj] {10.3847/1538-4357/abd8ce}, \href {https://ui.adsabs.harvard.edu/abs/2021ApJ...909...38D} {909, 38}

\bibitem[\protect\citeauthoryear{{Dere}, {Landi}, {Mason}, {Monsignori Fossi}  \& {Young}}{{Dere} et~al.}{1997}]{dere1997}
{Dere} K.~P.,  {Landi} E.,  {Mason} H.~E.,  {Monsignori Fossi} B.~C.,   {Young} P.~R.,  1997, \mn@doi [\aaps] {10.1051/aas:1997368}, \href {https://ui.adsabs.harvard.edu/abs/1997A&AS..125..149D} {125, 149}

\bibitem[\protect\citeauthoryear{{Fang}, {Buote}, {Bullock}  \& {Ma}}{{Fang} et~al.}{2015}]{fang2015}
{Fang} T.,  {Buote} D.,  {Bullock} J.,   {Ma} R.,  2015, \mn@doi [\apjs] {10.1088/0067-0049/217/2/21}, \href {https://ui.adsabs.harvard.edu/abs/2015ApJS..217...21F} {217, 21}

\bibitem[\protect\citeauthoryear{{Gatuzz} \& {Churazov}}{{Gatuzz} \& {Churazov}}{2018}]{gautzz2018}
{Gatuzz} E.,  {Churazov} E.,  2018, \mn@doi [\mnras] {10.1093/mnras/stx2776}, \href {https://ui.adsabs.harvard.edu/abs/2018MNRAS.474..696G} {474, 696}

\bibitem[\protect\citeauthoryear{{Gupta}, {Mathur}, {Krongold}, {Nicastro}  \& {Galeazzi}}{{Gupta} et~al.}{2012}]{gupta2012}
{Gupta} A.,  {Mathur} S.,  {Krongold} Y.,  {Nicastro} F.,   {Galeazzi} M.,  2012, \mn@doi [\apjl] {10.1088/2041-8205/756/1/L8}, \href {https://ui.adsabs.harvard.edu/abs/2012ApJ...756L...8G} {756, L8}

\bibitem[\protect\citeauthoryear{{Gupta}, {Mathur}, {Galeazzi}  \& {Krongold}}{{Gupta} et~al.}{2014}]{gupta2014}
{Gupta} A.,  {Mathur} S.,  {Galeazzi} M.,   {Krongold} Y.,  2014, \mn@doi [\apss] {10.1007/s10509-014-1958-z}, \href {https://ui.adsabs.harvard.edu/abs/2014Ap&SS.352..775G} {352, 775}

\bibitem[\protect\citeauthoryear{{Gupta}, {Mathur}  \& {Krongold}}{{Gupta} et~al.}{2017}]{gupta2017}
{Gupta} A.,  {Mathur} S.,   {Krongold} Y.,  2017, \mn@doi [\apj] {10.3847/1538-4357/836/2/243}, \href {https://ui.adsabs.harvard.edu/abs/2017ApJ...836..243G} {836, 243}

\bibitem[\protect\citeauthoryear{{Gupta}, {Kingsbury}, {Mathur}, {Das}, {Galeazzi}, {Krongold}  \& {Nicastro}}{{Gupta} et~al.}{2021}]{gupta2021}
{Gupta} A.,  {Kingsbury} J.,  {Mathur} S.,  {Das} S.,  {Galeazzi} M.,  {Krongold} Y.,   {Nicastro} F.,  2021, \mn@doi [\apj] {10.3847/1538-4357/abdbb6}, \href {https://ui.adsabs.harvard.edu/abs/2021ApJ...909..164G} {909, 164}

\bibitem[\protect\citeauthoryear{{Gupta}, {Mathur}, {Kingsbury}, {Das}  \& {Krongold}}{{Gupta} et~al.}{2023}]{gupta2023}
{Gupta} A.,  {Mathur} S.,  {Kingsbury} J.,  {Das} S.,   {Krongold} Y.,  2023, \mn@doi [Nature Astronomy] {10.1038/s41550-023-01963-5}, \href {https://ui.adsabs.harvard.edu/abs/2023NatAs.tmp...91G} {}

\bibitem[\protect\citeauthoryear{{Henley} \& {Shelton}}{{Henley} \& {Shelton}}{2013}]{henley2013}
{Henley} D.~B.,  {Shelton} R.~L.,  2013, \mn@doi [\apj] {10.1088/0004-637X/773/2/92}, 773, 92

\bibitem[\protect\citeauthoryear{{Henley}, {Shelton}, {Kwak}, {Joung}  \& {Mac Low}}{{Henley} et~al.}{2010}]{henley2010}
{Henley} D.~B.,  {Shelton} R.~L.,  {Kwak} K.,  {Joung} M.~R.,   {Mac Low} M.-M.,  2010, \mn@doi [\apj] {10.1088/0004-637X/723/1/935}, \href {https://ui.adsabs.harvard.edu/abs/2010ApJ...723..935H} {723, 935}

\bibitem[\protect\citeauthoryear{{Kaaret} et~al.,}{{Kaaret} et~al.}{2020}]{kaaret2020}
{Kaaret} P.,  et~al., 2020, \mn@doi [Nature Astronomy] {10.1038/s41550-020-01215-w}, \href {https://ui.adsabs.harvard.edu/abs/2020NatAs...4.1072K} {4, 1072}

\bibitem[\protect\citeauthoryear{{Kaastra} et~al.,}{{Kaastra} et~al.}{2018}]{kaastra2018}
{Kaastra} J.~S.,  et~al., 2018, \mn@doi [\aap] {10.1051/0004-6361/201832629}, \href {https://ui.adsabs.harvard.edu/abs/2018A&A...619A.112K} {619, A112}

\bibitem[\protect\citeauthoryear{{Krongold}, {Nicastro}, {Brickhouse}, {Elvis}, {Liedahl}  \& {Mathur}}{{Krongold} et~al.}{2003}]{krongold2003}
{Krongold} Y.,  {Nicastro} F.,  {Brickhouse} N.~S.,  {Elvis} M.,  {Liedahl} D.~A.,   {Mathur} S.,  2003, \mn@doi [\apj] {10.1086/378639}, \href {https://ui.adsabs.harvard.edu/abs/2003ApJ...597..832K} {597, 832}

\bibitem[\protect\citeauthoryear{Lara-DI, Mathur, Krongold, Das  \& Gupta}{Lara-DI et~al.}{2023}]{lara2023}
Lara-DI A.,  Mathur S.,  Krongold Y.,  Das S.,   Gupta A.,  2023, \mn@doi [The Astrophysical Journal] {10.3847/1538-4357/acbf40}, 946, 55

\bibitem[\protect\citeauthoryear{{Mathur}}{{Mathur}}{2022}]{mathur2022}
{Mathur} S.,  2022, in , Handbook of X-ray and Gamma-ray Astrophysics. Edited by Cosimo Bambi and Andrea Santangelo.
p.~59, \mn@doi{10.1007/978-981-16-4544-0_112-1}

\bibitem[\protect\citeauthoryear{{Nicastro}, {Senatore}, {Krongold}, {Mathur}  \& {Elvis}}{{Nicastro} et~al.}{2016}]{nicastro2016}
{Nicastro} F.,  {Senatore} F.,  {Krongold} Y.,  {Mathur} S.,   {Elvis} M.,  2016, \mn@doi [\apjl] {10.3847/2041-8205/828/1/L12}, \href {https://ui.adsabs.harvard.edu/abs/2016ApJ...828L..12N} {828, L12}

\bibitem[\protect\citeauthoryear{{Oppenheimer} et~al.,}{{Oppenheimer} et~al.}{2016}]{oppenheimer2016}
{Oppenheimer} B.~D.,  et~al., 2016, \mn@doi [\mnras] {10.1093/mnras/stw1066}, \href {https://ui.adsabs.harvard.edu/abs/2016MNRAS.460.2157O} {460, 2157}

\bibitem[\protect\citeauthoryear{{Salvato} et~al.,}{{Salvato} et~al.}{2018}]{rosatcat2018}
{Salvato} M.,  et~al., 2018, \mn@doi [\mnras] {10.1093/mnras/stx2651}, \href {https://ui.adsabs.harvard.edu/abs/2018MNRAS.473.4937S} {473, 4937}

\bibitem[\protect\citeauthoryear{{Snowden}, {Freyberg}, {Kuntz}  \& {Sanders}}{{Snowden} et~al.}{2000}]{snowden2000}
{Snowden} S.~L.,  {Freyberg} M.~J.,  {Kuntz} K.~D.,   {Sanders} W.~T.,  2000, \mn@doi [\apjs] {10.1086/313378}, \href {https://ui.adsabs.harvard.edu/abs/2000ApJS..128..171S} {128, 171}

\bibitem[\protect\citeauthoryear{{Tumlinson}, {Peeples}  \& {Werk}}{{Tumlinson} et~al.}{2017}]{tumlinson2017}
{Tumlinson} J.,  {Peeples} M.~S.,   {Werk} J.~K.,  2017, \mn@doi [\araa] {10.1146/annurev-astro-091916-055240}, \href {https://ui.adsabs.harvard.edu/abs/2017ARA&A..55..389T} {55, 389}

\bibitem[\protect\citeauthoryear{{Vijayan} \& {Li}}{{Vijayan} \& {Li}}{2022}]{vijayan2022}
{Vijayan} A.,  {Li} M.,  2022, \mn@doi [\mnras] {10.1093/mnras/stab3413}, \href {https://ui.adsabs.harvard.edu/abs/2022MNRAS.510..568V} {510, 568}

\bibitem[\protect\citeauthoryear{{Williams} et~al.,}{{Williams} et~al.}{2005}]{williams2005}
{Williams} R.~J.,  et~al., 2005, \mn@doi [\apj] {10.1086/431343}, \href {https://ui.adsabs.harvard.edu/abs/2005ApJ...631..856W} {631, 856}

\makeatother
\end{thebibliography}

%\begin{thebibliography}{99}
%\end{thebibliography}

%%%%%%%%%%%%%%%%%%%%%%%%%%%%%%%%%%%%%%%%%%%%%%%%%%

%%%%%%%%%%%%%%%%% APPENDICES %%%%%%%%%%%%%%%%%%%%%

\appendix

\section{PHASE Predictions}
Our non-solar PHASE model predicts the presence of many absorption lines, most of which are outside the wavelength range of Chandra or require a higher S/N than available with the current data. We provide a list of those predictions with the expected EW and N and their associate uncertainties. We only include those in the wavelength range of XRISM, as the Resolve instrument will provide the best opportunity to observe these lines.

The quoted column densities are the totals for each ion.
They have not been calculated with Equation \ref{curve-of-growth}, as in Table \ref{abs-lines}, but are simply an output of PHASE.

\begin{table*}
    \setlength\extrarowheight{4pt}
    \centering
    \caption{Predicted equivalent widths and column densities for all absorption lines in our non-solar 2-component PHASE model within the wavelength range of XRISM/Resolve. The quoted EW is the total EW for each of the lines, the sum of both phase contributions. N$_\text{tot}$ represents the total column density of the ion. Many of these lines should be detectable with XRISM when NGC\,3783 is in a high flux state.}
    \begin{tabular}{c@{\hskip 0.3in}c@{\hskip 0.25in}c@{\hskip 0.75in}cc}
        \hline
        % \hline
        % \multicolumn{5}{c}{PHASE$_1$}\\
        \hline
        Element & Ion & $\lambda$ ({\AA}) & EW (m{\AA}) & N$_\text{tot}$ $\left(\times 10^{15} \text{ cm}^{-2}\right)$ \\
        \hline
        \multirow{5}{1em}{C} & \ion{C}{V} & 40.267 & 29.14$^{+6.13}_{-6.37}$ & \multirow{3}{4em}{37.10$^{+23.80}_{-18.70}$} \\
        & \ion{C}{V} & 34.973 & 12.39$^{+1.50}_{-2.23}$ & \\
        & \ion{C}{V} & 33.426 & 8.47$^{+1.52}_{-2.47}$ & \\
        \cline{2-5}
        & \ion{C}{VI} & 33.736 & 11.34$^{+1.96}_{-2.90}$ & \multirow{2}{4em}{7.74$^{+4.96}_{-3.91}$}\\
        & \ion{C}{VI} & 28.466 & 3.35$^{+1.38}_{-1.46}$ &  \\
        \hline
        \multirow{2}{1em}{N} & \ion{N}{VI} & 28.787 & 13.56$^{+2.68}_{-3.11}$ & \multirow{2}{4em}{11.70$^{+7.50}_{-5.91}$} \\
        & \ion{N}{VI} & 24.898 & 5.09$^{+1.28}_{-1.80}$ & \\
        \hline
        \multirow{5}{1em}{O} & \ion{O}{VII} & 21.601 & 25.83$^{+7.16}_{-7.36}$ & \multirow{4}{4em}{94.10$^{+60.40}_{-47.55}$}\\
        & \ion{O}{VII} & 18.627 & 6.93$^{+1.21}_{-1.31}$ & \\
        & \ion{O}{VII} & 17.768 & 4.76$^{+0.82}_{-1.22}$ & \\
        & \ion{O}{VII} & 17.396 & 3.34$^{+0.81}_{-1.11}$ & \\
        \cline{2-5}
        & \ion{O}{VIII} & 18.969 & 5.41$^{+2.25}_{-2.34}$ & 5.43$^{+3.60}_{-2.74}$\\
        \hline
        \multirow{3}{1.1em}{Ne} & \ion{Ne}{IX} & 13.447 & 12.48$^{+3.36}_{-4.35}$ & \multirow{2}{4em}{20.51$^{+5.18}_{-4.05}$}  \\
        & \ion{Ne}{IX} & 11.544 & 3.05$^{+1.40}_{-2.55}$ &  \\
        \cline{2-5}
        & \ion{Ne}{X} & 12.134 & 4.29$^{+1.54}_{-1.75}$ & 11.70$^{+7.70}_{-5.94}$ \\
        \hline
        \multirow{1}{1em}{Ar} & \ion{Ar}{IX} & 35.024 & 2.24$^{+1.05}_{-1.03}$ & \multirow{1}{4em}{0.46$^{+0.29}_{-0.23}$} \\
        \hline
        \multirow{1}{1em}{Ca} &  \ion{Ca}{XI} & 30.471 & 2.38$^{+1.32}_{-1.15}$ &  0.12$^{+0.07}_{-0.06}$\\
        \hline
        \multirow{1}{1em}{Fe} &  \ion{Fe}{XVII} & 15.014 & 3.14$^{+1.73}_{-1.49}$ & 0.66$^{+0.43}_{-0.33}$ \\
        \hline
        
    \end{tabular}
    \label{phase-predictions}
\end{table*}

\bsp	% typesetting comment
\label{lastpage}
\end{document}